\documentclass[12pt]{article}
\usepackage{amssymb,amsmath,amsfonts,amsthm,amstext,amscd,array}
\usepackage{mathrsfs}
\usepackage{hyperref}
\usepackage{pdfsync}
\usepackage{bbm}
\usepackage{bm}
\usepackage[arrow,matrix,curve]{xy}
\usepackage{bbding}
\usepackage{wasysym}

\usepackage{amsfonts}
\usepackage{booktabs}
\usepackage{siunitx}

\usepackage{epsf}
\usepackage{epsfig}
\usepackage{wrapfig}
\input{epsf}

\newcommand{\eq}{\begin{equation}}
\newcommand{\eqend}{\end{equation}}
\newcommand{\eqa}{\begin{eqnarray}}
\newcommand{\nonueqa}{\begin{eqnarray*}}
\newcommand{\eqaend}{\end{eqnarray}}
\newcommand{\nonueqaend}{\end{eqnarray*}}

\newcommand{\bma}[1]{\begin{array}{#1}}
\newcommand{\ema}{\end{array}}
\newcommand{\bc}{\begin{center}}
\newcommand{\ec}{\end{center}}

\newif\ifold             \oldtrue

\def\be{\begin{equation}}
\def\ee{\end{equation}}
\def\bea{\begin{eqnarray}}
\def\eea{\end{eqnarray}}
\def\bd{\begin{displaymath}}
\def\ed{\end{displaymath}}

\newcommand{\beq}{\begin{eqnarray}}
\newcommand{\eeq}{\end{eqnarray}}

\makeatletter
\newdimen\normalarrayskip              
\newdimen\minarrayskip                 
\normalarrayskip\baselineskip
\minarrayskip\jot
\newif\ifold             \oldtrue            
\def\arraymode{\ifold\relax\else\displaystyle\fi} 
\def\@arrayskip{\ifold\baselineskip\z@\lineskip\z@
     \else
     \baselineskip\minarrayskip\lineskip2\minarrayskip\fi}
\def\@arrayclassz{\ifcase \@lastchclass \@acolampacol \or
\@ampacol \or \or \or \@addamp \or
   \@acolampacol \or \@firstampfalse \@acol \fi
\edef\@preamble{\@preamble
  \ifcase \@chnum
     \hfil$\relax\arraymode\@sharp$\hfil
     \or $\relax\arraymode\@sharp$\hfil
     \or \hfil$\relax\arraymode\@sharp$\fi}}
\def\@array[#1]#2{\setbox\@arstrutbox=\hbox{\vrule
     height\arraystretch \ht\strutbox
     depth\arraystretch \dp\strutbox
     width\z@}\@mkpream{#2}\edef\@preamble{\halign \noexpand\@halignto
\bgroup \tabskip\z@ \@arstrut \@preamble \tabskip\z@ \cr}%
\let\@startpbox\@@startpbox \let\@endpbox\@@endpbox
  \if #1t\vtop \else \if#1b\vbox \else \vcenter \fi\fi
  \bgroup \let\par\relax
  \let\@sharp##\let\protect\relax
  \@arrayskip\@preamble}
\makeatother

\newcommand{\p}{\partial}

\def\be{\beta}

\newtheorem{proposition}[equation]{Proposition}
\newtheorem{definition}[equation]{Definition}

\theoremstyle{definition}

\usepackage{tempora}
\usepackage{color,xcolor}
\usepackage{manfnt}

\def\p{\vskip2ex\hskip5ex}
\def\ddo{\end{document}}
\usepackage{hyperref}
\hypersetup{colorlinks,%
citecolor=black,%
filecolor=black,%
linkcolor=black,%
urlcolor=black}

\newcommand{\nc}{\newcommand}
\nc{\lb}{\llbracket}
\nc{\rb}{\rrbracket}
\nc{\gl}{\llbracket}
\nc{\gr}{\rrbracket}

\begin{document}

\title{\Large\bf  Poisson gauge theory}

\author{{ Vladislav G. Kupriyanov}\thanks{Email: \ {\tt
    vladislav.kupriyanov@gmail.com}}\\
    \\
{\it Centro de Matem\'atica, Computa\c{c}\~{a}o e
Cogni\c{c}\~{a}o}\\{\it Universidade Federal do ABC, Santo Andr\'e, SP, 
Brazil}\\ and {\it 
Tomsk State University, Tomsk, Russia }  }

\maketitle
\p
\begin{abstract}
The Poisson gauge algebra is a semi-classical limit of complete non-commutative gauge algebra. In the present work we formulate the Poisson gauge theory which is a dynamical field theoretical model having the Poisson gauge algebra as a corresponding algebra of gauge symmetries. The proposed model is designed to investigate the semi-classical features of the full non-commutative gauge theory with coordinate dependent non-commutativity $\Theta^{ab}(x)$, especially whose with a non-constant rank. We derive the expression for the covariant derivative of matter field. The commutator relation for the covariant derivatives defines the Poisson field strength which is covariant under the Poisson gauge transformations and reproduces the standard $U(1)$ field strength in the commutative limit. We derive the corresponding Bianchi identities. The field equations for the gauge and the matter fields are obtained from the gauge invariant action. We consider different examples of linear in coordinates Poisson structures $\Theta^{ab}(x)$, as well as non-linear ones, and obtain explicit expressions for all proposed constructions. Our model is unique up to invertible field redefinitions and coordinate transformations.
\end{abstract}

\section{Introduction}

The consistent definition of non-commutative gauge theory is an important step for understanding of low-energy physics of D-branes in general backgrounds of string theory \cite{SW}-\cite{HullSzabo}. This problem was extensively investigated during the last decades. We mention here the main approaches discussed in the literature: the covariant coordinates approach \cite{Wess,Wess1}, the twist approach \cite{Dimitrijevic}-\cite{twist1}, the L$_\infty$-bootstrap approach \cite{BBKL} and also recently proposed approach based on the combination of the L$_\infty$ algebras and the twist \cite{Ciric}. Nevertheless the structure of the non-commutative field theory still requires the better understanding, especially in case when the non-commutativity $\Theta^{ab}(x)$ is not constant. In the previous work \cite{KS21} in collaboration with Richard Szabo we formulated a novel approach to the construction of algebra of non-commutative gauge symmetries in the semi-classical limit, based on the symplectic embeddings of (almost)-Poisson gauge structure. The aim of the present research is to construct the dynamical sector of the corresponding theory. 

The associative non-commutativity of space-time is usually introduced in the theory by substituting the standard pointwise multiplication of fields $f\cdot g$ on some manifold $M$ with the star multiplication,
\begin{equation}\label{star}
f\star g=f\cdot g+\frac{i\hbar}{2}\{f,g\}+{\cal O}(\hbar^2)\,,
\end{equation}
 defined as an associative deformation of the pointwise product along the given Poisson structure,
 \begin{equation}\label{PB}
\{x^i,x^j\}=\alpha\,\Theta^{ij}(x)\,,
\end{equation}
where $\alpha$ is a small parameter which we will refere to as the non-commutativity parameter. The non-commutativity field $\Theta^{ij}(x)$ is considered to be an external field. The higher order in $\hbar$ contributions to (\ref{star}) are defined from the condition of the associativity, $(f\star g)\star h=f\star(g\star h)$, and can be constructed according to the Formality theorem \cite{Kontsevich}, or also the polydifferential approach \cite{KV08}.

The non-commutative $U(1)$ gauge transformations $\delta^{NC}_fA_a$ are defined as transformations satisfying the following two conditions: they should close the algebra,
\begin{equation}
[\delta^{NC}_f,\delta^{NC}_g]A_a=\delta^{NC}_{-i[f,g]_\star}A_a\,,\label{ga1}
\end{equation}
with, $[f,g]_\star=f\star g-g\star f$, and should reproduce the standard $U(1)$ gauge transformations $\delta^0_fA_a=\partial_af$, in the commutative limit.  Reminding that the Poisson bracket is the semi-classical limit of the star commutator, $\{f,g\}=\lim_{\hbar\to0}[f,g]_\star/i\hbar$, we define the Poisson gauge transformations $\delta_fA_a$, following \cite{KS21}, as the semi-classical limit of the full non-commutative $U(1)$ gauge transformations. They should close the algebra,
\begin{equation}
[\delta_f,\delta_g]A_a=\delta_{\{f,g\}} A_a\,,\label{ga}
\end{equation}
called Poisson gauge algebra, and reproduce the standard $U(1)$ gauge transformations in the commutative limit, $\lim_{\alpha\to0}\delta_fA_a=\partial_af$.

If $\Theta^{ij}$ is constant, one may easily see that the expression, $\delta_fA_a=\partial_af+\{f,A_a\}$, satisfies (\ref{ga}). However, for non-constant $\Theta^{ij}(x)$ the standard Leibniz rule with respect to the partial derivative is violated, $\partial_a\{f,g\}\neq\{\partial_af,g\}+\{f,\partial_ag\}$, therefore the same expression will not close the algebra (\ref{ga}) anymore. To overcome this problem one has to modify the expression for the gauge transformations introducing the corrections proportional to the derivatives of the non-commutativity $\partial_a\Theta^{ij}(x)$ which would compensate the violation of the Leibniz rule. The problem was solved in \cite{Kupriyanov:2019ezf} using the L$_\infty$-bootstrap approach to the non-commutative gauge theories \cite{BBKL}. We stress that L$_\infty$-formalism is a powerful tool for the construction of perturbative order by order in $\alpha$ expressions for the consistent non-commutative deformations of gauge theories \cite{Kup27}. Though, to get an explicit all order expressions normally one needs to invoke additional considerations. 

The approach to the solution of this problem proposed in \cite{KS21} is based on the symplectic embeddings of Poisson manifolds \cite{SE1,SE2} and is especially good for obtaining explicit form of the deformed constructions. The problem with violation of the Leibniz rule for the original Poisson bracket (\ref{PB}) can be solved in an extended space. To each coordinate $x^i$ we introduce a conjugate variable $p_i$, in such a way that the corresponding Poisson brackets,
\begin{equation}\label{PB1}
\{x^i,p_j\}=\gamma^i_j(x,p)\,,\qquad \{p_i,p_j\}=0\,,
\end{equation}
should satisfy the Jacobi identity. In our construction we will need the vanishing bracket between $p$-variables, $\{p_i,p_j\}=0$, while the matrix, $\gamma^i_j(x,p)=\delta^i_j-\mbox{$\frac{\alpha}{2}$}\partial_j\Theta^{ik}p_k+{\cal O}(\alpha^2)$, defining the Poisson bracket $\{x^i,p_j\}$ will be constructed in Sec. 2. For constant $\Theta^{ij}$ this matrix is constant, $\gamma^i_j(x,p)=\delta^i_j$, so, $\{f(x),p_i\}=\partial_if(x)$, i.e., the Poisson bracket between the function $f(x)$ on $M$ and the auxiliary variable $p_i$ is just a partial derivative of this function. In case if $\Theta^{ij}(x)$ is  not constant the expression for $\gamma^i_j(x,p)$ is more complicated, however the action of the operator $\{\,\cdot\,,p_i\}$ which we will call `twisted' derivative on functions $f(x)$ is similar to that of the partial derivatives $\partial_i$. First of all because the Jacobi identity for the Poisson brackets (\ref{PB1}) and the fact that $\{p_i,p_j\}=0$ imply that these operators comute, $\{\{f(x),p_i\},p_j\}=\{\{f(x),p_j\},p_i\}$. Second, because the `twisted' derivative satisfies the Leibniz rule,
\begin{equation}\label{Lr}
\{\{f(x),g(x)\},p_i\}=\{\{f(x),p_i\},g(x)\}+\{f(x),\{g(x),p_i\}\}\,,
\end{equation}
which also follows from the Jacobi identity. However, the price to pay is that the expression $\{f(x),p_i\}$ depends also on the auxiliary non-physical $p$-variables. It turns out that the auxiliary variables can be eliminated in the consistent way by introducing the constraints, $p_a=A_a(x)$\footnote{Note that in \cite{KS18} the symplectic embeddings were used to construct the consistent Hamiltonian description of the electrically charged particle in the field of magnetic monopole distributions. However in that case the elimination the auxiliary $p$-variables was not possible.}. In the Sec. 3 we will prove that the gauge transformations defined by,
\begin{equation}\label{gtA}
\delta_f A_a=\gamma^l_a(A)\,\partial_lf(x)+\{A_a(x),f(x)\}\,,
\end{equation}
where, $\gamma^l_a(A):=\gamma^l_a(x,p)|_{p_a=A_a(x)}$, close the algebra (\ref{ga}) and reproduce the standard $U(1)$ gauge transformations, $\delta^0_fA_a=\partial_af$, in the commutative limit.

The new results of the present research are related to the consistent definition of the dynamical part of field-theoretical model having the Poisson gauge algebra (\ref{ga}) as the corresponding algebra of gauge symmetries. Working in the formalism of the symplectic embeddings we introduce the matter field $\psi$ by postulating the corresponding gauge transformation as, $\delta_f\psi=\{\psi,f\}$. In the Sec. 4 we construct the gauge covariant derivative ${\cal D}_a(\psi)$ satisfying two key requirements: it transforms covariantly under the gauge transformation, $\delta_f{\cal D}_a(\psi)=\{{\cal D}_a(\psi),f\}$, and reproduces the standard partial derivative in the commutative limit, ${\cal D}_a(\psi)\to\partial_a\psi$, as $\alpha\to0$. In the Sec. 5 it will be shown that the commutator of two covariant derivatives defines the Poisson field strength ${\cal F}_{ab}$ which also transforms covariantly, $\delta_f{\cal F}_{ab}=\{{\cal F}_{ab},f\}$, and reproduces the standard $U(1)$ field strength in the commutative limit, $\lim_{\alpha\to0}{\cal F}_{ab}=\partial_aA_b-\partial_bA_a$. We also define the corresponding Bianchi identity in the Sec. 5.1. The main relations are resumed by the following table, which for simplicity is given for the case of linear Poisson structures, $\Theta^{ij}(x)=f^{ij}_kx^k$,
\begin{center}
\begin{tabular}{S|  |S} \toprule
    {\mbox{Object}}  & {Identity}  \\ \midrule \midrule
  {${\cal D}_a(\psi)=\rho_a^i(A)\left(\gamma_i^l(A)\,\partial_l\psi+\{A_a,\psi\}\right)$}  & {$\left[{\cal D}_a,{\cal D}_b\right]=\{{\cal F}_{ab},\,\cdot\,\}+\left({\cal F}_{ad}\,\Lambda_b{}^{de}-{\cal F}_{bd}\,\Lambda_a{}^{de}\right){\cal D}_e$} \\  \midrule
    {${\cal F}_{ab}=R_{ab}{}^{ij}(A)(\gamma_i^l(A)\,\partial_l A_j+\{A_i,A_j\})$} & {${\cal D}_a\left({\cal F}_{bc}\right)-{\cal F}_{ad}\,\Lambda_b{}^{de}\,{\cal F}_{ec}+\mbox{cycl.}(abc)=0$}    \\ \bottomrule
\end{tabular}
\end{center}
where,
\begin{equation*}
R_{ab}{}^{cd}=\frac12\left(\rho^c_a\rho^d_b-\rho^c_b\rho^d_a\right)\,,\qquad\Lambda_b{}^{de}=\left(\rho^{-1}\right)_j^d\left(\partial^j_A\rho_b^m-\partial^m_A\rho_b^j\right)\left(\rho^{-1}\right)_m^e\,,
\end{equation*}
with, $\partial^j_A=\partial/\partial A_j$, and the matrix $\rho_a^i(A)$ is defined by, $\rho_a^i(A):=\rho^i_a(x,p)|_{p_a=A_a(x)}$, where $\rho^i_a(x,p)$ should satisfy the equation,
\begin{equation}\label{rhoi}
\{f(x),\rho_a^i(x,p)\}+\rho_a^b(x,p)\,\partial_p^ i\{f(x),p_b\}=0\,, \qquad \forall f(x)\,.
\end{equation}
For the non-linear Poisson structures $\Theta^{ij}(x)$ there may appear additional contributions in the identities in the right hand side of the table which are given in the Sec. 5.

For arbitrary non-commutativity $\Theta^{ij}(x)$ we provide the recurrence relations for the construction of the matrices, $\gamma_a^l(A)$ and $\rho_a^l(A)$  which are the building blocs of our construction. For some particular choices of the non-commutative spaces, like the rotationally invariant NC space \cite{Kup13}-\cite{Gere} described by the $su(2)$-like Lie Poisson structure, $\Theta^{ij}_\varepsilon(x)=2\alpha\varepsilon^{ijk}x_k$, or the $\kappa$-Minkowski space \cite{kappa}-\cite{kappa3} we obtain explicit all-order expressions for $\gamma_a^l(A)$ and $\rho_a^l(A)$ in the Sections 2 and 4 correspondingly. In the Sec. 6 we use the gauge covariant objects ${\cal D}_a(\psi)$ and ${\cal F}_{ab}$ to construct the gauge invariant action and derive the field equations for the gauge field $A_a$ and the matter field $\psi$. In case of the rotationally invariant non-commutative space \cite{Kup13} the field equations in the pure gauge sector read, ${\cal D}^{\varepsilon}_a\left({\cal F}^{ab}_\varepsilon\right)=0$. We conclude with the final remarks and discussion in the Sec. 7 and provide the useful for the calculation formulas in the Appendix.

\section{Symplectic embeddings of Poisson manifold}

In this section we will summarize the necessary ingredients from the symplectic geometry that we will use throughout the paper. All precise mathematical definitions and proofs are given in \cite{KS21}, here our aim is to recollect them  and expose on a physics friendly language. The problem of the construction of the symplectic embedding for the given Poisson structure (\ref{PB}) formulated in the introduction consists basically in finding the matrix $\gamma_j^i(x,p)$ which defines the Poisson bracket $\{x^i,p_j\}$ in such a way that the complete algebra of Poisson brackets (\ref{PB}) and (\ref{PB1}) should satisfy the Jacobi identity. 

The Jacobi identity involving the original coordinates only, $\{x^i,\{x^j,x^k\}\}+\mbox{cycl.}=0$, is satisfied automatically since $\Theta^{ij}(x)$ is a Poisson bi-vector. The Jacobi identity with two original coordinates and one $p$-variable,
$
\{x^i,\{x^j,p_k\}\}+\mbox{cycl.}=0\,,
$
implies an equation\footnote{Note that here we use different notations from \cite{KS21}, the matrix $\gamma_a^k$ here corresponds to $\delta_a^k+\alpha\,\gamma_a^k$ used in \cite{KS21}.},
\begin{equation}
\gamma^l_b\,\partial^b_p\gamma_a^k-\gamma^k_b\,\partial^b_p\gamma^l_a+\alpha\,\Theta^{lm}\,\partial_m\gamma_a^k-\alpha\,\Theta^{km}\,\partial_m\gamma_a^l-\alpha\,\gamma^m_a\,\partial_m\Theta^{lk}=0\,,\label{eq2}
\end{equation}
on the function $\gamma^i_j(x,p)$ in terms of given Poisson structure $\Theta^{ij}(x)$.
The Jacobi identity with one $x$ and two $p$--variables, $\{x^i,\{p_j,p_k\}\}+\mbox{cycl.}=0$, is also satisfied automatically. It follows from the fact that, $\{p_i,p_j\}=0$, and antisymmetry of $\{x^j,p_k\}$. And finally the Jacobi identity involving the $p$-variables only, $\{p_i,\{p_j,p_k\}\}+\mbox{cycl.}=0$, is trivially satisfied.

So, the matrix $\gamma_j^i(x,p)$ is defined as a solution of the equation (\ref{eq2}) with the condition, $\gamma_j^i(x,p)|_{\alpha=0}=\delta^i_j$, to guarantee that the complete algebra of the Poisson brackets (\ref{PB}) and (\ref{PB1}) is a deformation in $\alpha$ of the canonical Poisson brackets, i.e., forms the symplectic algebra. Up to the second order in $\alpha$ the solution reads,
\begin{eqnarray}
\gamma^k_a(x,p)=\sum_{n=0}\,\gamma^{k(n)}_a&=&\delta^k_a-\frac{\alpha}{2}\, \partial_a \Theta^{kb} p_b\label{gps}\\
&&-\frac{\alpha^2}{12}\left(2\,\Theta^{cm}\partial_a\partial_m\Theta^{bk}+\partial_a\Theta^{bm}\partial_m\Theta^{kc}\right)p_bp_c+{\cal O}(\alpha^3)\,.\notag
\end{eqnarray}
The recurrence relations for the construction of the matrix $\gamma^k_a(x,p)$ in any order in $\alpha$ are given in \cite{Kup21}.

\subsection{Arbitrariness}

We note that the symplectic embedding is not unique. The arbitrariness is described by the invertible transformations of the variables $(x,p)$ which leave the Poisson brackets between the coordinates and coordinates $\{x^i,x^j\}$, as well as, momenta and momenta $\{p_i,p_j\}$ unchanged, while change the brackets between the coordinates and momenta $\{x^i,p_j\}$. Making the transformation which change only the $p$-variables, $\phi:\,p\to\tilde p$, with $\phi(p)|_{\alpha=0}=p$ and leave $x$-variables unchanged we do not change neither the bracket (\ref{PB}), nor the $\{p_i,p_j\}=0$, while for the bracket between the coordinates and momenta one finds,
\begin{eqnarray}\label{tildegamma}
\tilde \gamma^i_j(x,\tilde p)=\{x^i,\tilde p_j(p)\}_{p=p(\tilde p)}=\gamma^i_k(x,p)\,\partial^k_p\tilde p_j(p)|_{p=p(\tilde p)}\,.
\end{eqnarray}
The new set of Poisson brackets,
\begin{eqnarray}
\{x^i,x^j\}&=&\alpha\,\Theta^{ij}(x)\,,\\
\{x^i,\tilde p_j\}&=&\tilde \gamma^i_j(x,\tilde p)\,,\qquad \{\tilde p_i,\tilde p\}=0\,,\notag
\end{eqnarray}
also represents the symplectic embedding of the Poisson structure (\ref{PB}) since $\tilde \gamma^i_j(x,\tilde p)|_{\alpha=0}=\delta^i_j$.

\subsection{Symplectic embeddings of Lie-Poisson structures}

The Kirillov-Kostant Poisson bracket, also called sometimes the Lie-Poisson bracket, is defined in \cite{Alekseev} as,
\begin{equation}\label{e1}
\{x^i,x^j\}=f^{ij}_k\,x^k\,,
\end{equation}
where $f^{ij}_k$ are the structure constants of a Lie algebra. The solution for the equation (\ref{eq2}) in this case is given by \cite{KS21,Gutt,Meljanac},
 \begin{equation}\label{e4}
   \gamma^i_j(p)=\delta^i_j-\frac12\,f^{ij_1}_j\,p_{j_1}+\mathcal{X}^i_j\left(-M/{2}\right)\,,
\end{equation}
where, \begin{equation}\label{e4a}
M^{i}_l={f}^{ij_1}_k {f}^{kj_2}_lp_{j_1}p_{j_2}\,,
\end{equation}
 and $\mathcal{X}^i_l\left(-M/{2}\right)$ is a matrix valued function with,
\begin{equation}\label{e5}
\mathcal{X}(t)=\sqrt{\frac{t}{2}}\cot\sqrt{\frac{t}{2}}-1=\sum_{n=1}^\infty\frac{(-2)^n\,B_{2n}\,t^{n}}{(2n)!}\,,
\end{equation}
with $B_n$ being Bernoulli numbers. We stress that the matrix $\gamma^i_j(p)$ defined in (\ref{e4}) does not depend on $x$-variables.

Let us consider an exemple of the $su(2)$-like Lie-Poisson algebra, 
\begin{equation}
\{x^k,x^l\}=2\,\alpha\,{\varepsilon^{kl}}_m\, x^m\,,\label{su2}
\end{equation}
physically corresponding to the rotationally invariant non-commutative space. In (\ref{su2}) $\varepsilon^{klm}$ is the Levi-Civita symbol and the factor of $2$ is just a matter of convenience. We use the Kronecker delta to raise and lower  indices,
and  summation under the repeated indices is understood. In this case the matrix $M_\varepsilon$ appearing as the argument of the third term in the r.h.s. of (\ref{e4}) is given by,
\begin{equation}\label{e6}
[M_\varepsilon]^i_l=4\,\alpha^2\left(p^i\,p_l-\delta^i_l\,p^2\right)\,,
\end{equation}
where, $p^2=p_mp^m$. This matrix is diagonalizable, it can be written as, $M_\varepsilon=S\cdot D\cdot S^{-1}$, where $D$ is the diagonal matrix with the eigenvalues of $M_\epsilon$,  $\lambda_1=0$, $\lambda_2=\lambda_3=-4\alpha^2p^2$, on the diagonal and the matrix $S$ is constructed from the corresponding eigenvectors. Therefore following \cite{KV15} we write,
\begin{eqnarray}\label{e7}
\mathcal{X}^i_j\left(-M_\varepsilon/{2}\right)&=&\left[S\cdot \rm{diag}\left[ \mathcal{X}\left(-\lambda_1/{2}\right),\mathcal{X}\left(-\lambda_2/{2}\right),\mathcal{X}\left(-\lambda_3/{2}\right)\right]\cdot S^{-1}\right]^{i}_j\\
&=& \alpha^2\left(\delta^i_j\,p^2- p^ ip_j\right)\chi(\alpha^2\,p^2)\,,\notag
\end{eqnarray}
where,
\begin{equation}\label{chi}
\chi(t)=t^{-1}\,\mathcal{X}(2t)=\frac1t\,\left(\sqrt{t}\cot\sqrt{t}-1\right)\,.
\end{equation}
We conclude that,
\begin{equation}
[\gamma_\varepsilon]^k_a(p)=\left[1+\alpha^2p^2\chi\left(\alpha^2p^2\right)\right]\delta^k_a-\alpha^2\chi\left(\alpha^2p^2\right)p_ap^k-\alpha\,\varepsilon_a{}^{kl}p_l\,,\label{gammasu2}
\end{equation}
Note that the explicit form of the function $\chi(t)$ was obtained from the general solution (\ref{e4}) diagonalizing the matrix $\mathcal{X}^i_j\left(-M_\varepsilon/{2}\right)$. Alternatively one may check that the expression (\ref{gammasu2}) satisfies the equation (\ref{eq2}) only if $\chi(t)$ obeys the ODE,
\begin{equation}\label{w21}
2\,t\,\chi^\prime+3\,\chi+t\,\chi^2+1=0\,,\qquad \chi(0)=-\frac13\,,
\end{equation}
whose solution is given by (\ref{chi}).

The second particular exemple we would like to discuss here is the $\kappa$-Minkowski space \cite{kappa}-\cite{kappa3} yielding the Kirillov-Kostant structure,
\begin{equation}
\{x^k,x^l\}=2\left(a^i\,x^j-a^j\,x^i\right)\,,\label{kappa}
\end{equation}
with $a^i$ being constants. For this Poisson structure the simplest solution for $\gamma^k_a(p)$ reads \cite{KKV},
\begin{equation}
[\gamma_\kappa]^k_a(p)= \left[\sqrt{1+(a\cdot p)^2}+(a\cdot p)\right]\delta^k_a -a^k\,p_a\,.\label{gammakappa}
\end{equation}

\subsection{Change of coordinates}

If two different Poisson structures $\Theta^{ij}(x)$ and $\tilde\Theta^{ij}(\tilde x)$ are related by the invertible coordinate transformation, $\upsilon:\,x\to\tilde x$, i.e.,
\begin{equation}\label{g1}
\tilde\Theta^{ij}(\tilde x)=\{\tilde x^i(x),\tilde x^j(x)\}_{x=x(\tilde x)}=\partial_k\tilde x^i\,\Theta^{kl}(x)\,\partial_l\tilde x^j|_{x=x(\tilde x)}\,,
\end{equation}
and the symplectic embedding $\gamma^i_j(x,p)$ of the first one is known, then the symplectic embedding of the second one can be constructed according to,
\begin{equation}\label{g2}
\tilde \gamma^i_j(\tilde x,p)=\{\tilde x^i(x),p_j\}_{x=x(\tilde x)}=\partial_k\tilde x^i\, \gamma^k_j(x,p)|_{x=x(\tilde x)}\,.
\end{equation}
Thus formally having in hand a symplectic embedding of a given Poisson structure we may generate symplectic embeddings of the whole family of the related Poisson structures. At the same time we have to be careful with the applications of this construction since the physical observables should not depend on the change of coordinates.

Let us consider an example. We start with the symplectic embedding (\ref{gammasu2}) of the $su(2)$-like Lie-Poisson structure (\ref{su2}). The corresponding non-commutativity preserves rotational symmetry, see \cite{Kup13}. Let us discuss its generalization, namely the non-linear Poisson structure,
\begin{equation}
\{\tilde x^i,\tilde x^j\}=2\,\alpha\,{\varepsilon^{ij}}_m\, \tilde x^m\,f(\tilde x^2)\,,\label{su2f}
\end{equation}
where $f(\tilde x^2)$ is some given function. The non-commutativity (\ref{su2f}) also preserves rotations, however it does not grows on the infinity provided that $f(\tilde x^2)$ decreases faster then $1/\sqrt{\tilde x^2}$, when $\tilde x^2\to\infty$. In this case we have a kind of "local" rotationally invariant non-commutativity. The algebra (\ref{su2f}) can be obtained from (\ref{su2}) by the change of variables,
\begin{equation}\label{g3}
\tilde x^i=g(x^2)\,x^i\,,
\end{equation}
where $g(x^2)$ is a function to be determined. Note that, $\tilde x^2=g^2(x^2)\,x^2$. One calculates,
\begin{equation}\label{g4}
\{\tilde x^i,\tilde x^j\}=2\,\alpha\,{\varepsilon^{ij}}_m\, x^m\,g^2(x^2)=2\,\alpha\,{\varepsilon^{ij}}_m\, \tilde x^m\,g(x^2)\,.
\end{equation}
Comparing (\ref{g4}) to (\ref{su2f}) and taking into account (\ref{g3}) one finds,
\begin{equation}\label{g5}
f\left(g^2(x^2)\,x^2\right)=g(x^2)\,,
\end{equation}
which is a functional relation on $g$ provided that $f$ is some given function. Taking $f$ as the Gaussian function, $f(\tilde x^2)=\exp(-\alpha\tilde x^2/2)$, one obtains from (\ref{g5}),
\begin{equation}\label{g6}
g(x^2)=e^{-\frac{W(\alpha x^2)}{2}}\,,\qquad\Rightarrow\qquad \tilde x^i=e^{-\frac{W(\alpha x^2)}{2}}\,x^i\,,
\end{equation}
where $W(z)$ is the Lambert $W$ function satisfying the functional relation, $z=W(z)\,e^{W(z)}$. One may easily verify that in this case, $\alpha\tilde x^2=\exp(-W(\alpha x^2))\, \alpha x^2=W(\alpha x^2)$, and thus, $f(\tilde x^2)=\exp(-\alpha\tilde x^2/2)=\exp(-W(\alpha x^2)/2)=g(x^2)$, i.e., (\ref{g5}) holds true. The inverse transformation is,
\begin{equation}\label{g7}
x^i=\frac{1}{g(x^2)}\,\tilde x^i=e^{\frac{\alpha\tilde x^2}{2}}\,\tilde x^i\,.
\end{equation}
Now from (\ref{g2}), (\ref{g3}) and (\ref{g7}) we derive the symplectic embedding of (\ref{su2f}),
\begin{equation}
\tilde\gamma^i_j(\tilde x,p)=e^{-\frac{\alpha\tilde x^2}{2}}\left(\delta^i_k-\frac{\alpha\,\tilde x_k\,\tilde x^i}{1+\alpha\,\tilde x^2}\right)\gamma^k_j(p)\,,
\end{equation}
where the matrix $\gamma^k_j(p)$ was constructed in (\ref{gammasu2}) and we remind that we do not change $p$ variables.

\section{Poisson gauge transformations}

In this section we prove the statement formulated in the introduction regarding the closure condition (\ref{ga}) of the Poisson gauge transformations (\ref{gtA}).
First we observe that the (\ref{gtA}) can be written in a more convenient form as,
\begin{equation}\label{APhi}
\delta_f A_a=\{f(x),\Phi_a\}_{\Phi=0}\,,
\end{equation}
where,
\begin{equation}\label{Phi}
\Phi_a:=p_a-A_a(x)\,.
\end{equation}
Indeed, the expression $\{f(x),p_a\}_{p_a-A_a(x)=0}$ means that first we calculate the Poisson bracket $\{f(x),p_a\}$ according to (\ref{PB1}) and then substitute all $p$ with $A$, so,
\begin{equation*}
\{f(x),p_a\}_{\Phi=0}=\gamma^l_a(A)\,\partial_lf(x)\,.
\end{equation*}
The main property of the transformation (\ref{APhi}) if formulated by the following proposition.
{\proposition \label{pgt} The Poisson gauge transformations (\ref{APhi}) close the algebra (\ref{ga}).}
\proof We start with some preliminary definitions,
\begin{equation}
\delta_f F(A):=F\left(A+\delta_f A\right)-F(A)\,.
\end{equation}
In particular, 
\begin{equation}
\delta_g \left(\{A_a, f\}\right)=\{\delta_g A_a,f\}\,,
\end{equation}
and if $F(A)$ is a smooth function in $A$, then,
\begin{equation}
\delta_f F(A)=\left(\partial_A^b F\right)\,\delta_f A_b\,,
\end{equation}
where as before we use the notation, $\partial_A^b=\partial/\partial A_b$. The composition of two gauge transformations can be written,
\begin{eqnarray}
\delta_f\left(\delta_g A_a\right)&=&-\{g(x), \delta_f A_a\}_{\Phi=0}+\left(\partial^b_p\{g(x),\Phi_a\}\right)_{\Phi=0}\delta_f A_b\\
&=&-\{g(x), \{f(x),\Phi_a\}_{\Phi=0}\}_{\Phi=0}+\left(\partial^b_p\{g(x),\Phi_a\}\right)_{\Phi=0}\{f(x),\Phi_b\}_{\Phi=0}\,,\notag
\end{eqnarray}
since,
\begin{equation}
\partial_A^b\left(\{g(x),\Phi_a\}_{\Phi=0}\right)=\left(\partial_p^b \{g(x),\Phi_a\}\right)_{\Phi=0}\,.\label{a1}
\end{equation}
And thus we find,
\begin{eqnarray}\label{t6}
&&\delta_f\left(\delta_g A_a\right)-\delta_g\left(\delta_f A_a\right)=\\
&&-\{g(x), \{f(x),\Phi_a\}_{\Phi=0}\}_{\Phi=0}+\left(\partial^b_p\{g(x),\Phi_a\}\right)_{\Phi=0}\{f(x),\Phi_b\}_{\Phi=0}\notag\\
&&+\{f(x), \{g(x),\Phi_a\}_{\Phi=0}\}_{\Phi=0}-\left(\partial^b_p\{f(x),\Phi_a\}\right)_{\Phi=0}\{g(x),\Phi_b\}_{\Phi=0}\,.\notag
\end{eqnarray}
Using the relation (\ref{r3}) from the appendix in the right hand side of (\ref{t6}) we represent it as,
\begin{equation}
-\{g(x), \{f(x),\Phi_a\}\}_{\Phi=0}+\{f(x), \{g(x),\Phi_a\}\}_{\Phi=0}\,.
\end{equation}
Finally applying now the Jacobi identity we end up with,
\begin{equation}\label{t7}
\delta_f\left(\delta_g A_a\right)-\delta_g\left(\delta_f A_a\right)=\{\{f(x),g(x)\},\Phi_a\}_{\Phi=0}=\delta_{\{f,g\}}A_a\\.
\end{equation}
The latter means that the gauge variations (\ref{APhi}) close the Lie algebra (\ref{ga}). $\Box$

\subsection{Field redefinition} 

The form of the Poisson gauge transformation (\ref{APhi}) relies on the precise symplectic embedding $\gamma^i_a(x,p)$ of given Poisson structure (\ref{PB}). 
However, as it was discussed in Sec. 2.1 the symplectic embedding of (\ref{PB}) is not unique. Different embeddings are related by the invertible change of $p$-variables, $\phi:\,p\to\tilde p$, with $\phi(p)|_{\alpha=0}=p$. Having the Poisson gauge transformation (\ref{APhi}) which satisfy the relation (\ref{ga}), one may construct another gauge transformation corresponding to the new symplectic embedding $\tilde\gamma^i_a(x,\tilde p)$,
\begin{equation}\label{fr1}
\tilde\delta_f\tilde A_a=\tilde\gamma^i_a(\tilde A)\,\partial_if+\{f,\tilde A_a\}\,,
\end{equation}
which will close the same gauge algebra (\ref{ga}).

From the physical perspective one may expect that the invertible field redefinitions defined by, $\phi: A\to\tilde A$, and $\tilde  f=f$,
will leave our construction invariant. We define the new gauge transformations $\tilde\delta_f$ by setting,
\begin{equation}
\tilde\delta_f:=\phi\circ\delta_f\circ \phi^{-1}\,,
\end{equation}
or in the other words,
\begin{equation}
\tilde\delta_f \tilde A_a:=\left.\delta_f \tilde A_a(A)\right|_{A(\tilde A)}=\left.\partial^k_A \tilde A_a(A) \,\delta_f A_k\right|_{A(\tilde A)}\,.
\end{equation}
Taking into account (\ref{tildegamma}) we see that it is exactly the expression (\ref{fr1}), meaning that the arbitrariness in our construction corresponds to the invertible field redefinition.

Since the field redefinition is invertible the gauge orbits are mapped onto the gauge orbits and the Seiberg-Witten condition,
\begin{equation}
\tilde A\left(A+\delta_ fA\right)=\tilde A(A)+\tilde\delta_f\tilde A(A)\,,
\end{equation}
is trivially satisfied up to the linear order in $f$. Moreover,
\begin{equation}
\left[\tilde\delta_f,\tilde\delta_g\right]=\tilde\delta_f\circ\tilde\delta_g-\tilde\delta_g\circ\tilde\delta_f=\phi\circ\left(\delta_f\circ\delta_g-\delta_g\circ\delta_f\right)\circ\phi^{-1}=\tilde\delta_{\{f,g\}}\,,
\end{equation}
meaning that the gauge algebra remains the same, as expected. In \cite{BBKT} it was shown that the Seiberg-Witten maps correspond to L$_\infty$-quasi-isomorphisms which describe the arbitrariness in the definition of the related L$_\infty$ algebra in the L$_\infty$-bootstrap approach \cite{BBKL}.

\section{Covariant derivative}

We start with the introduction of the matter fields $\psi$ and the definition of the non-commutative $U(1)$ gauge transformations $\delta^{NC}_f\psi$ which should close the same gauge algebra (\ref{ga1}) as the corresponding gauge fields $A_a$. In case pf the associative star products there are three possibilities to define such a gauge transformation. It can be taken to be left or right star multiplication, $\delta^{NC}_f\psi=if\star\psi$ or $\delta^{NC}_f\psi=i\psi\star f$, or even as the star commutator, $\delta^{NC}_f\psi=-i[\psi,f]_\star$. Only the third possibility is compatible with the semi-classical limit which is the subject of this research. That is why we define the Poisson gauge transformation of the matter field as,
\begin{equation}\label{w3}
\delta_f\psi=\{\psi,f\}\,.
\end{equation}
The Jacobi identity implies that such a determined gauge transformations close the same algebra as (\ref{ga}), i.e.,
\begin{equation}
[\delta_f,\delta_g]\psi=\delta_{\{f,g\}}\psi\,.\label{w4}
\end{equation}

We note that in the commutative limit the Poisson gauge variation of the mater field vanishes, $\lim_{\alpha\to0}\delta_f\psi=0$,  meaning that in this limit the matter field $\psi$ is 'electrically neutral' and does not interact with the gauge field $A_a$. The interaction between the gauge field and such a defined matter field is caused by the non-commutativity. At the moment we do not see a clear physical meaning of such an interaction. The field $\psi$ appears in our construction more like an auxiliary object needed for the consistent definition of the covariant derivative, which in turn is a central object of this work. As we will see in the next Section, the commutator of the covariant derivatives will produce the gauge covariant field strength for the gauge field. Also this object is essential for the derivation of the corresponding Bianchi identity and the equations of motion.

Our aim now is to construct the covariant derivative ${\cal D}_a(\psi)$, the object satisfying two main requirements.  
\begin{itemize}
\item It should transform covariantly under the Poisson gauge transformation $\delta_f$ defined in (\ref{APhi}) and (\ref{w3}), i.e.,
\begin{equation}\label{w5}
\delta_f\left({\cal D}_a(\psi)\right)=\{{\cal D}_a(\psi),f\}\,.
\end{equation}
This property will be essential for the construction of the gauge covariant Lagrangian and the gauge invariant action for the Poisson gauge theory.
\item  In the commutative limit it should reproduce the standard partial derivative,
\begin{equation}\label{w5a}
\lim_{\alpha\to0}{\cal D}_a(\psi)=\partial_a\psi\,,
\end{equation}
 since the interaction between the matter field and the gauge field disappears when $\alpha\to 0$.
\end{itemize}

 The answer is given by the following proposition.

{\proposition \label{pcd} The operator,
\begin{equation}\label{w6}
{\cal D}_a(\psi)=\rho_a^i(A)\,\{\psi,\Phi_i\}_{\Phi=0}\,,
\end{equation}
satisfies the above two conditions if the matrix, $\rho_a^i(A):=\rho^i_a(x,p)_{\Phi=0}$, and $\rho^i_a(x,p)$ obeys the equation,
\begin{equation}\label{rho}
\{f(x),\rho_a^i(x,p)\}+\rho_a^b(x,p)\,\partial_p^ i\{f(x),p_b\}=0\,, \qquad \forall f(x)\,.
\end{equation}}
\proof
Let us calculate,
\begin{eqnarray}\label{w7}
&&\delta_f\left(\rho_a^i(A)\,\{\psi,\Phi_i\}_{\Phi=0}\right)=\\
&&\partial_A^b\rho_a^i(A)\,\{f,\Phi_b\}_{\Phi=0}\,\{\psi,\Phi_i\}_{\Phi=0}+\rho_a^i(A)\,\{\{\psi,f\},\Phi_i\}_{\Phi=0}+\notag\\
&&\rho_a^i(A)\,(\partial_p^ b\{\psi,\Phi_i\})_{\Phi=0}\,\{f,\Phi_b\}_{\Phi=0}-\rho_a^i(A)\,\{\psi,\{f,\Phi_i\}_{\Phi=0}\}_{\Phi=0}\,.\notag
\end{eqnarray}
Observe that,
\begin{equation}\label{w7a}
\partial_A^b\rho_a^i(A)\,\{f,\Phi_b\}_{\Phi=0}=\{f,\rho_a^i(p)-\rho_a^i(A)\}_{\Phi=0}\,,
\end{equation}
and by (\ref{r3}) from the appendix,
\begin{equation}
\{\psi,\{f,\Phi_i\}_{\Phi=0}\}_{\Phi=0}=\{\psi,\{f,\Phi_i\}\}_{\Phi=0}-(\partial_p^ b\{f,\Phi_i\})_{\Phi=0}\,\{\psi,\Phi_b\}_{\Phi=0}\,.
\end{equation}
So the right hand side of (\ref{w7}) becomes,
\begin{eqnarray}
&&\{f,\rho_a^i(p)\}_{\Phi=0}\,\{\psi,\Phi_i\}_{\Phi=0}+\{\rho_a^i(A),f\}_{\Phi=0}\,\{\psi,\Phi_i\}_{\Phi=0}+\\
&&\rho_a^i(A)\,\{\{\psi,f\},\Phi_i\}_{\Phi=0}+\rho_a^i(A)\,(\partial_p^ b\{\psi,\Phi_i\})_{\Phi=0}\,\{f,\Phi_b\}_{\Phi=0}+\notag\\
&&\rho_a^i(A)\,(\partial_p^ b\{f,\Phi_i\})_{\Phi=0}\,\{\psi,\Phi_b\}_{\Phi=0}-\rho_a^i(A)\,\{\psi,\{f,\Phi_i\}\}_{\Phi=0}\,.\notag
\end{eqnarray}
Using the Jacobi identity we rewrite it as,
\begin{eqnarray}\label{w8}
&&\left[\{f,\rho_a^i(p)\}_{\Phi=0}+\rho_a^b(A)\,(\partial_p^ i\{f,\Phi_b\})_{\Phi=0}\right]\{\psi,\Phi_i\}_{\Phi=0}+\\
&&\{\rho_a^i(A),f\}_{\Phi=0}\,\{\psi,\Phi_i\}_{\Phi=0}+\rho_a^i(A)\left[(\partial_p^ b\{\psi,\Phi_i\})_{\Phi=0}\,\{f,\Phi_b\}_{\Phi=0}+\{\{\varphi,\Phi_i\},f\}_{\Phi=0}\right]\,.\notag
\end{eqnarray}
Taking into account that by (\ref{r3}),
\begin{equation}
(\partial_p^ b\{\varphi,\Phi_i\})_{\Phi=0}\,\{f,\Phi_b\}_{\Phi=0}+\{\{\varphi,\Phi_i\},f\}_{\Phi=0}=\{\{\varphi,\Phi_i\}_{\Phi=0},f\}_{\Phi=0}\,,
\end{equation}
the second line in (\ref{w8}) becomes,
\begin{equation}
\{\rho_a^i(A)\,\{\varphi,\Phi_i\}_{\Phi=0},f\}_{\Phi=0}=\{{\cal D}_a(\varphi),f\}\,.
\end{equation}
Therefore the relation (\ref{w5}) holds true if the first line of (\ref{w8}) vanishes, i.e.,
\begin{equation}
\{f(x),\rho_a^i(x,p)\}+\rho_a^b(x,p)\,\partial_p^ i\{f(x),\Phi_b\}=0\,, \qquad \forall f(x)\,.
\end{equation}
Since $\{A_b(x),f(x)\}$ does not depend on $p$ one arrives at (\ref{rho}). The statement regarding the commutative limit, $\alpha\to 0$, follows from the fact that in this limit, $\{\psi,\Phi_i\}\to\partial_i\psi$, and $\rho_a^i(x,p)\to\delta_a^i$. $\Box$

In local coordinates the equation (\ref{rho}) on the function $\rho_a^i(x,p)$ can be written as,
\begin{equation}\label{w10}
\gamma^j_b\,\partial^b_p\, \rho_a^i+\rho_a^b\,\partial^i_p\gamma^j_b+\alpha\,\Theta^{jb}\,\partial_b\rho_a^i=0\,,
\end{equation}
where $\Theta^{jb}(x)$ is a given Poisson structure and $\gamma^j_b(x,p)$ represents its symplectic embedding constructed in the Sec. 2.
For the arbitrary non-commutativity parameter $\Theta^{ab}(x)$ a solution of the equation (\ref{w10}) can be found in form of the perturbative series,
\begin{eqnarray}
\rho^i_a(x,p)=\sum_{n=0}^\infty\,\rho^{i(n)}_a&=&\delta^i_a-\frac{\alpha}{2}\, \partial_a \Theta^{ib} p_b+\label{w11}\\
&&\frac{\alpha^2}{6}\left(2\,\Theta^{cm}\partial_a\partial_m\Theta^{ib}-\partial_a\Theta^{bm}\partial_m\Theta^{ic}\right)p_bp_c+{\cal O}(\alpha^3)\,.\notag
\end{eqnarray}
For some specific choices of  non-commutativity, one may also discuss the convergence of such series and exhibit a closed expressions.

\subsection{Lie-Poisson structures and generalizations}

First we observe that for the linear Poisson bi-vectors (\ref{e1}) there is no explicit coordinate dependance of the functions $\gamma_a^i(p)$ and $\rho_a^i(p)$. These functions depend only on $p$-variables $p_a$. The equation (\ref{w10}) becomes,
\begin{equation}\label{w11}
\gamma^j_l\,\partial^l_p\, \rho_a^i+\rho_a^l\,\partial^i_p\gamma^j_l=0\,,
\end{equation}
 where $\gamma^j_l(p)$ is given by (\ref{e4}). In what follows we will discuss the solution of the eq. (\ref{w11}) for two particular examples of Lie-Poisson structures.

\subsubsection{$su(2)$-like Poisson structure}
 
Let us consider now the particular case of $su(2)$-like Lie-Poisson structure (\ref{su2}). The perturbative calculations indicate the ansatz, 
 \begin{equation}\label{w12}
[ \rho_\varepsilon]_a^i=\sigma\left(\alpha^2p^2\right)\delta_a^i+\alpha^2\,\tau\left(\alpha^2p^2\right)p^ip_a+\alpha\,\zeta\left(\alpha^2p^2\right)\varepsilon_a{}^{ik}p_k\,,
\end{equation}
with the initial condition, $\sigma(0)=1$, to match (\ref{w5a}). Using this ansatz in the equation (\ref{w11}) the latter becomes,
\begin{eqnarray}\label{w13}
&&\alpha\left[2\,\sigma^\prime+\sigma\,\chi+\zeta\right]\delta^i_ap^j+\alpha\left[\tau+\zeta-\sigma\,\chi\right] \delta^{ij}p_a+\\
&&\alpha\left[\left(1+\alpha^2\,p^2\,\chi\right)\tau+2\,\sigma\left(\chi+\alpha^2\,p^2\,\chi^\prime\right)\right]\delta^j_ap^i+\alpha\left[\left(1+\alpha^2\,p^2\,\chi\right)\zeta+\sigma\right]\varepsilon_a{}^{ij}+\notag\\
&&\alpha^3\left[2\,\zeta^\prime-2\,\zeta\,\chi\right]p^j\varepsilon_a{}^{ik}p_k+\alpha^3\left[2\,\zeta \left(\chi+\alpha^2\,p^2\,\chi^\prime\right)-\tau \right]p^i\varepsilon_a^{jk}p_k+\notag\\
&&\alpha^4\left[2\,\tau^\prime-2\,\sigma\,\chi^\prime-\tau\,\chi\right]p^i\,p^j\,p_a=0\,.\notag
\end{eqnarray}
Thus the equation (\ref{w11}) results in the system of seven equations on coefficient functions $\sigma(\alpha^2p^2)$, $\tau(\alpha^2p^2)$ and $\zeta(\alpha^2p^2)$:
\begin{eqnarray}
&&2\,\sigma^\prime+\sigma\,\chi+\zeta=0\,,\label{w16}\\
&&\tau+\zeta-\sigma\,\chi=0\,,\label{w17}\\
&&\left(1+\alpha^2\,p^2\,\chi\right)\tau+2\,\sigma\left(\chi+\alpha^2\,p^2\,\chi^\prime\right)=0\,,\label{w18}\\
&&\left(1+\alpha^2\,p^2\,\chi\right)\zeta+\sigma=0\,,\label{w19}\\
&&2\,\zeta^\prime-2\,\zeta\,\chi=0\,,\label{w20}\\
&&2\,\zeta \left(\chi+\alpha^2\,p^2\,\chi^\prime\right)-\tau=0\,\label{w20a} \\
&&2\,\tau^\prime-2\,\sigma\,\chi^\prime-\tau=0\,.\label{w20b}
\end{eqnarray}
However they are not all independent. The eq. (\ref{w18}) is just a consequence of (\ref{w19}) and (\ref{w20a}). The eq. (\ref{w17}) is satisfied due (\ref{w19}), (\ref{w20a}) and the equation on the function $\chi$ (\ref{w21}). The eq. holds true as a consequence of (\ref{w19}), (\ref{w20}) and (\ref{w21}). The eq. (\ref{w20b}) is satisfied due (\ref{w19}), (\ref{w20}), (\ref{w20a}) and (\ref{w21}).

In fact we end up with only three independent equations on $\sigma(A)$, $\tau(A)$ and $\zeta(A)$: (\ref{w19}), (\ref{w20}) and (\ref{w20a}). Let us first discuss the initial conditions. Taking into account the initial condition, $\sigma(0)=1$, one finds from (\ref{w19}) that, $\zeta(0)=-1$. Taking into account that $\chi(0)=-1/3$, the equation (\ref{w17}) then gives the initial condition for the function $\tau$ namely, $\tau(0)=2/3$. The solution of the equation (\ref{w20}) with this initial condition reads,
\begin{equation}\label{w22}
\zeta(t)=-\frac{\sin^2\sqrt{t}}{t}\,.
\end{equation}
The equation (\ref{w19}) gives,
\begin{equation}\label{w23}
\sigma(t)=\frac{\sin2\sqrt{t}}{2\sqrt{t}}\,.
\end{equation}
And from the equation (\ref{w20a}) one finds,
\begin{equation}\label{w24}
\tau(t)=-t^{-1}\left(\frac{\sin2\sqrt{t}}{2\sqrt{t}}-1\right)\,.
\end{equation}
We conclude that the ansatz (\ref{w12}) with the functions $\sigma$, $\tau$ and $\zeta$ defined in (\ref{w23}), (\ref{w24}) and (\ref{w22}) correspondingly solve the equation (\ref{w11}).

An explicit form of the matrix, $\rho_a^i(A)=\rho^i_a(x,p)_{p=A}$, reads,
 \begin{equation}\label{rhosu2}
 [ \rho_\varepsilon]_a^i(A)=\delta_a^i+\alpha\,\varepsilon_a{}^{ik}A_k\,\zeta\left(\alpha^2\,A^2\right)- \alpha^2\left(\delta^i_a\,A^2- A^ iA_a\right)\tau(\alpha^2\,A^2)\,.
\end{equation}
And thus we have obtained an explicit all orders expression for the gauge covariant derivative defined in (\ref{w6}). It satisfies the gauge covariance condition (\ref{w5}) and reproduces in the commutative limit the standard partial derivative (\ref{w5a}). 

\subsubsection{$\kappa$-Minkowski}

For the $\kappa$-Minkowski Poisson structure (\ref{kappa}) the matrix $[\gamma_\kappa]^j_l(p)$ is given by (\ref{gammakappa}). Following the same strategy as in the previous particular example we write the anzats which follows from the perturbative calculations,
 \begin{equation}\label{rhokappa}
[ \rho_\kappa]_a^i(p)=\delta_a^i\,\xi(z)+a^i\,p_a\,\eta(z)\,,
\end{equation}
where, $z=a\cdot p$, and $\xi(0)=1$ to guarantee the correct commutative limit. Substituting this anzats in (\ref{w11}) one finds,
\begin{eqnarray}
\left((\phi -z)\,\xi^\prime-\xi \right)\delta_a^i\,a^j+\left(\phi\,\eta+\xi\,\phi^\prime\right)\delta_a^j\,a^i+\left((\phi-z)\,\eta^\prime-2\eta+\phi^\prime\,\eta\right)a^i\,a^j\,p_a=0\,,
\end{eqnarray}
where, $\phi(z)=\sqrt{1+z^2}+z$, according to (\ref{gammakappa}). Consequently one arrives at the system of three equations on the coefficient functions $\xi$ and $\eta$:
\begin{eqnarray}\label{eqxe}
&&(\phi -z)\,\xi^\prime=\xi\,,\\
&&\phi\,\eta+\xi\,\phi^\prime=0\,,\notag\\
&&(\phi-z)\,\eta^\prime-2\eta+\phi^\prime\,\eta=0\,.\notag
\end{eqnarray}
The solution of the first one with the initial condition, $\xi(0)=1$, reads,
\begin{equation}
\xi(z)=\sqrt{1+z^2}+z\,.
\end{equation}
Then from the second equation one finds,
\begin{equation}
\eta(z)=-\frac{\sqrt{1+z^2}+z}{\sqrt{1+z^2}}\,.
\end{equation}
The third of the equations (\ref{eqxe}) just becomes an identity.

\subsubsection{Change of coordinates}

Under the change of coordinates, $\upsilon:\,x\to \tilde x$, described in Sec. 2.4 the solution of the equation (\ref{rho}) transforms like,
\begin{equation}\label{rhotilde}
\upsilon:\, \rho_a^i( x,p)\to\tilde \rho_a^i(\tilde x,p)= \rho_a^i( x,p)|_{x=x(\tilde x)}\,.
\end{equation}
For Lie-Poisson structures it does not change, since $\rho_a^i( p)$ does not depend on coordinates $x^i$.

\section{ Field strength}

Let us first discuss the commutator relation of the covariant derivatives ${\cal D}_a(\psi)$ introduced in the previous section.
{\proposition \label{cr} The commutator relation for the covariant derivatives defined in the proposition \ref{pcd} reads,
\begin{eqnarray}\label{cr1}
\left[{\cal D}_a,{\cal D}_b\right]=\{{\cal F}_{ab},\,\cdot\,\}+\left({\cal F}_{ad}\,\Lambda_b{}^{de}-{\cal F}_{bd}\,\Lambda_a{}^{de}+{\cal K}_{ab}{}^e-{\cal K}_{ba}{}^e\right){\cal D}_e\,
\end{eqnarray}
where,
\begin{eqnarray}\label{F}
{\cal F}_{ab}:&=&\rho_a^i(A)\,\rho_b^j(A)\,\{\Phi_i,\Phi_j\}_{\Phi=0}\\
&=&\rho_a^i(A)\,\rho_b^j(A)\left(\gamma_i^l(A)\,\partial_lA_j-\gamma_j^l(A)\,\partial_lA_i+\{A_i,A_j\}\right)\,,\notag\\
\Lambda_b{}^{de}(A)&=&\left(\rho^{-1}\right)_j^d\left(\partial^j_A\rho_b^m(A)-\partial^m_A\rho_b^j(A)\right)\left(\rho^{-1}\right)_m^e\,,\label{Lambda}\\
{\cal K}_{ab}{}^e(A)&=&\rho_a^i(A)\,\gamma^m_i(A)\,\left(\partial_m\rho_b^j(x,p)\right)_{\Phi=0}\left(\rho^{-1}\right)_j^e\,.\label{Kappa}
\end{eqnarray}
\proof The proof is straightforward, we start writing, 
\begin{eqnarray}\label{f2}
\left[{\cal D}_a,{\cal D}_b\right](\psi)&=&\left(\rho_a^i(A)\,\{\rho_b^j(A),\Phi_i\}_{\Phi=0}-\rho_b^i(A)\,\{\rho_a^j(A),\Phi_i\}_{\Phi=0}\right)\{\psi,\Phi_j\}_{\Phi=0}+\\
&&\rho_a^i(A)\,\rho_b^j(A)\left(\{\{\psi,\Phi_j\}_{\Phi=0},\Phi_i\}_{\Phi=0}-\{\{\psi,\Phi_i\}_{\Phi=0},\Phi_j\}_{\Phi=0}\right)\,.\notag
\end{eqnarray}
We use the formula (\ref{r5}) from the appendix in the second line of the right hand side of the above expression to rewrite it as,
\begin{eqnarray}\label{f3}
&&\left(\rho_a^i(A)\,\{\rho_b^j(A),\Phi_i\}_{\Phi=0}-\rho_b^i(A)\,\{\rho_a^j(A),\Phi_i\}_{\Phi=0}\right)\{\psi,\Phi_j\}_{\Phi=0}+\\
&&\rho_a^i(A)\,\rho_b^j(A)\left(\{\{\psi,\Phi_j\},\Phi_i\}_{\Phi=0}-\{\{\psi,\Phi_i\},\Phi_j\}_{\Phi=0}\right)\notag\\
&&-\rho_a^i(A)\,\rho_b^j(A)\left[\left(\partial_p^m\{\psi,\Phi_j\}\right)_{\Phi=0}\,\{\Phi_m,\Phi_i\}_{\Phi=0}-\left(\partial_p^m\{\psi,\Phi_i\}\right)_{\Phi=0}\,\{\Phi_m,\Phi_j\}_{\Phi=0}\right]\,.\notag
\end{eqnarray}
Employing the Jacobi identity in the second line and the equation (\ref{rho}) in the third line of (\ref{f3}) one represents it as,
\begin{eqnarray}\label{f4}
&&\left(\rho_a^i(A)\,\{\rho_b^j(A),\Phi_i\}_{\Phi=0}-\rho_b^i(A)\,\{\rho_a^j(A),\Phi_i\}_{\Phi=0}\right)\{\psi,\Phi_j\}_{\Phi=0}+\\
&&\rho_a^i(A)\,\rho_b^j(A)\{\{\Phi_i,\Phi_j\},\psi\}_{\Phi=0}+\notag\\
&&\left(\rho_a^i(A)\{\rho_b^j(p),\psi\}_{\Phi=0}- \rho_b^i(A)\{\rho_a^j(p),\psi\}_{\Phi=0}\right)\{\Phi_i,\Phi_j\}_{\Phi=0}\notag
\end{eqnarray}
Applying one more time the eq. (\ref{r5}) in the second line of (\ref{f4}) we reorganize it as, 
\begin{eqnarray}\label{f5}
&&\left(\rho_a^i(A)\,\{\rho_b^j(A),\Phi_i\}_{\Phi=0}-\rho_b^i(A)\,\{\rho_a^j(A),\Phi_i\}_{\Phi=0}\right)\{\psi,\Phi_j\}_{\Phi=0}+\\
&&\{\rho_a^i(A)\,\rho_b^j(A)\,\{\Phi_i,\Phi_j\}_{\Phi=0},\psi\}_{\Phi=0}-\rho_a^i(A)\,\rho_b^j(A)\,\left(\partial_p^m\{\Phi_i,\Phi_j\}\right)_{\Phi=0}\,\{\Phi_m,\psi\}_{\Phi=0}+\notag\\
&&\left(\rho_a^i(A)\{\rho_b^j(p)-\rho_b^j(A),\psi\}_{\Phi=0}- \rho_b^i(A)\{\rho_a^j(p)-\rho_a^j(A),\psi\}_{\Phi=0}\right)\{\Phi_i,\Phi_j\}_{\Phi=0}\,.\notag
\end{eqnarray}
We use the equation (\ref{rho}) in the second term of the second line and also (\ref{w7a}) in the last line rewriting (\ref{f5}) as,
\begin{eqnarray}\label{f6}
&&\rho_a^i(A)\left(\{\rho_b^j(A),\Phi_i\}_{\Phi=0}-\{A_i,\rho_b^j(p)\}_{\Phi=0}\right)\{\psi,\Phi_j\}_{\Phi=0}\\
&&-\rho_b^i(A)\left(\{\rho_a^j(A),\Phi_i\}_{\Phi=0}-\{A_i,\rho_a^j(p)\}_{\Phi=0}\right)\{\psi,\Phi_j\}_{\Phi=0}+\notag\\
&&\{\rho_a^i(A)\,\rho_b^j(A)\,\{\Phi_i,\Phi_j\}_{\Phi=0},\psi\}_{\Phi=0}+\notag\\
&&\left(\rho_a^i(A)\,\partial^m_A\rho_b^j(A)\{\Phi_m,\psi\}_{\Phi=0}- \rho_b^i(A)\,\partial^m_A\rho_a^j(A)\{\Phi_m,\psi\}_{\Phi=0}\right)\{\Phi_i,\Phi_j\}_{\Phi=0}\,.\notag
\end{eqnarray}

Let us calculate separately,
\begin{eqnarray}\label{f7}
&&\{\rho_b^j(A),\Phi_i\}_{\Phi=0}-\{A_i,\rho_b^j(p)\}_{\Phi=0}=\\
&&\partial^m_A\rho_b^j(A)\left(\{A_m,p_i\}_{\Phi=0}-\{A_i,p_m\}_{\Phi=0}-\{A_m,A_i\}\right)+\left(\partial_m\rho_b^j(x,p)\right)_{\Phi=0}\,\gamma^m_i(A)=\notag\\
&&-\partial^m_A\rho_b^j(A)\,\{\Phi_m,\Phi_i\}_{\Phi=0}+\left(\partial_m\rho_b^j(x,p)\right)_{\Phi=0}\,\gamma^m_i(A)\,.\notag
\end{eqnarray}
After the use the relation (\ref{f7}) in the expression (\ref{f6}) the latter becomes,
\begin{eqnarray}\label{f8}
&&\{\rho_a^i(A)\,\rho_b^j(A)\,\{\Phi_i,\Phi_j\}_{\Phi=0},\psi\}_{\Phi=0}\\
&&+\rho_a^i(A)\left(\partial^j_A\rho_b^m(A)-\partial^m_A\rho_b^j(A)\right)\{\Phi_i,\Phi_j\}_{\Phi=0}\,\{\psi,\Phi_m\}_{\Phi=0}\notag\\
&&-\rho_b^i(A)\left(\partial^j_A\rho_a^m(A)-\partial^m_A\rho_a^j(A)\right)\{\Phi_i,\Phi_j\}_{\Phi=0}\,\{\psi,\Phi_m\}_{\Phi=0}\notag\\
&&+\rho_a^i(A)\,\gamma^m_i(A)\,\partial_m\rho_b^j(A)\,\{\psi,\Phi_j\}_{\Phi=0}-\rho_b^i(A)\,\gamma^m_i(A)\,\partial_m\rho_a^j(A)\,\{\psi,\Phi_j\}_{\Phi=0}\,.\notag
\end{eqnarray}
Finally using the definitions (\ref{F}), (\ref{Lambda}) and (\ref{Kappa}) as well as the formula (\ref{w6}) we represent (\ref{f8}) as the right hand side of (\ref{cr1}) and thus prove the proposition \ref{cr}. $\Box$

Some comments are in order. First of all we remind that in case of Lie-Poisson structures which are the main example of the present work, the matrix $\rho_a^i(p)$ does not depend on coordinates. Therefore the tensor ${\cal K}_{ab}{}^{e}$ defined in (\ref{Kappa}) vanishes and the commutator relation becomes just,
\begin{eqnarray}\label{cr2}
\left[{\cal D}_a,{\cal D}_b\right]=\{{\cal F}_{ab},\,\cdot\,\}+\left({\cal F}_{ad}\,\Lambda_b{}^{de}-{\cal F}_{bd}\,\Lambda_a{}^{de}\right){\cal D}_e\,
\end{eqnarray}
For the canonical Poisson bracket with constant non-commutativity parameter $\Theta^{ij}$ the matrix, $\rho_a^i(A)=\delta_a^i$, is also constant according to (\ref{w11}). So, the tensor $\Lambda_b{}^{de}$ determined in (\ref{Lambda}) vanishes and the relation (\ref{cr2}) becomes the 'usual' one, $\left[{\cal D}_a,{\cal D}_b\right]=\{{\cal F}_{ab},\,\cdot\,\}$.

Second, in the commutative limit the Poisson bracket vanishes and $$\lim_{\alpha\to0}\rho_a^i(A)=\lim_{\alpha\to0}\gamma_a^i(A)=\delta_a^i\,,$$ so the quantity introduced in (\ref{F}) reproduces the $U(1)$ field strength when $\alpha\to0$,
\begin{equation}\label{clF}
\lim_{\alpha\to0}\,{\cal F}_{ab}=\partial_a\,A_b-\partial_b\,A_a\,.
\end{equation}}
Moreover, this object enjoys another important property formulated by the following,
{\proposition The quantity (\ref{F}) introduced in the proposition \ref{cr} transforms covariantly under the Poisson gauge transformation (\ref{APhi}), 
\begin{equation}\label{gcF}
\delta_f {\cal F}_{ab}=\{{\cal F}_{ab},f\}\,.
\end{equation}
\proof The proof of this proposition is analogous to the proposition \ref{pcd}. $\Box$

Because of the properties (\ref{clF}) and (\ref{gcF}) we will call the quantity ${\cal F}_{ab}$ defined in (\ref{F}) as the {\it Poisson field strength}.

In the previous works \cite{KV20,KKV} we have derived the quasi-classical limit of the non-commutative $U(1)$ field strength as the quantity, 
\begin{equation}
{\cal F}_{ab}=P_{ab}{}^{cd}\left(A\right)\,\partial_c A_d+R_{ab}{}^{cd}\left(A\right)\,\left\{A_c,A_d\right\}\,,\label{F1}
\end{equation}
which satisfies the same properties (\ref{clF}) and (\ref{gcF}). In (\ref{F1}), $P_{ab}{}^{cd}=2\,\gamma^c_lR_{ab}{}^{ld}$, and the coefficient function $R_{ab}{}^{cd}$ should satisfy the equation,
\begin{equation}
\gamma^k_l\,\partial^l_A R_{ab}{}^{cd}+\alpha\,\Theta^{kl}\,\partial_lR_{ab}{}^{cd}+R_{ab}{}^{cl}\,\partial^d_A\gamma^k_l+R_{ab}{}^{ld}\,\partial^c_A\gamma^k_l=0\,.\label{eqR}
\end{equation}
Comparing (\ref{F}) and (\ref{F1}) we see that,
\begin{equation}\label{Rrho}
R_{ab}{}^{cd}(A)=\frac12\left(\rho^c_a(A)\rho^d_b(A)-\rho^c_b(A)\rho^d_a(A)\right)\,.
\end{equation}
Substituting (\ref{Rrho}) in the l.h.s. of (\ref{eqR}) one finds,
\begin{eqnarray}\label{f9}
&&\frac12\left[\gamma^k_l\,\partial^l_A\, \rho_a^c+\rho_a^l\,\partial^c_A\gamma^k_l+\alpha\,\Theta^{kl}\,\partial_l\rho_a^c\right]\rho^d_b+\\
&&\frac12\,\rho^c_a\left[\gamma^k_l\,\partial^l_A\, \rho_b^d+\rho_b^l\,\partial^d_A\gamma^k_l+\alpha\,\Theta^{kl}\,\partial_l\rho_b^d\right]-(a\leftrightarrow b)\,.\notag
\end{eqnarray}
Observing that in the square brackets we have the l.h.s. of the eq. (\ref{w10}) written in $A$ variables we conclude that if $\rho_a^c$ obeys the equation (\ref{rho}), the tensor $R_{ab}{}^{cd}(A)$ defined by the relation (\ref{Rrho}) satisfies the equation (\ref{eqR}). 

It is easy to see that substituting the expression (\ref{rhokappa}) for the matrix $[\rho_\kappa]_a^c$ corresponding to the $\kappa$-Minkowski Lie-Poisson structure (\ref{kappa}) in (\ref{Rrho}) one obtains exactly the formula for $R_{ab}^\kappa{}^{cd}$ from \cite{KKV}. The calculation in case of the $su(2)$-like Poisson bi-vector (\ref{su2}) is a bit less straightforward. We substitute $[\rho_\varepsilon]_a^c$ from (\ref{rhosu2}) in (\ref{Rrho}) and then use the formulas from the appendix of \cite{Kup27} to simplify it. Thus one recovers the expression for $R_{ab}^\varepsilon{}^{cd}$ from \cite{KV20}. And finally we note that if the Poisson structures $\Theta^{ij}(x)$ and $\tilde\Theta^{ij}(\tilde x)$ are related by the invertible coordinate transformation, $\upsilon:\,x\to\tilde x$, described in the Sec. 2.3, then the corresponding Poisson field strengths are also related by the same transformation, $\tilde{\cal F}_{ab}= {\cal F}_{ab}|_{x=x(\tilde x)}$, as can be seen from (\ref{g1}), (\ref{g2}), (\ref{rhotilde}) and (\ref{F}).

From the practical reasons the eq. (\ref{rho}) on $\rho_a^c$ is easier to solve than the eq. (\ref{eqR}) on $R_{ab}{}^{cd}$ because it has less components. However there is more fundamental significance of the covariant derivative (\ref{w6}), since it is essential for the definition of the Bianchi identity, the action principle and the equations of motion. It is therefore one may think of the matrices $\rho_a^c(A)$ and $\gamma_k^l(A)$ as a building blocs of the Poisson gauge theory.

\subsection{Bianchi identity}

{\proposition \label{bi} The field strength defined in (\ref{F}) satisfies the deformed Bianchi identity,
\begin{equation}\label{bi1}
{\cal D}_a\left({\cal F}_{bc}\right)-{\cal F}_{ad}\,\Lambda_b{}^{de}\,{\cal F}_{ec}-\Gamma_{ab}{}^e\,{\cal F}_{ec}+\Gamma_{cb}{}^e\,{\cal F}_{ea}+\mbox{cycl.}(abc)=0\,.
\end{equation}}
\proof Again the proof is straightforward. Using the definitions (\ref{w6}) and (\ref{F}) one calculates,
\begin{eqnarray}\label{b2}
&&{\cal D}_a\left({\cal F}_{bc}\right)+\mbox{cycl.}(abc)=\\
&&\rho_a^i(A)\left[\{ \rho_b^j(A),\Phi_i\}_{\Phi=0}\,\rho_c^k(A)+\rho_b^j(A)\,\{\rho_c^k(A),\Phi_i\}_{\Phi=0}\right]\{\Phi_j,\Phi_k\}_{\Phi=0}+\notag\\
&&\rho_a^i(A)\,\rho_b^j(A)\,\rho_c^(A)\left\{\{\Phi_j,\Phi_k\}_{\Phi=0},\Phi_i\right\}_{\Phi=0}+\mbox{cycl.}(abc)\,.\notag
\end{eqnarray}
In the last term on the right we use the formula (\ref{r6}) from the appendix and rewrite it as,
\begin{eqnarray}\label{b3}
&&\rho_a^i(A)\,\rho_b^j(A)\,\rho_c^k(A)\left\{\{\Phi_j,\Phi_k\}_{\Phi=0},\Phi_i\right\}_{\Phi=0}+\mbox{cycl.}(abc)=\\
&&\rho_a^i(A)\,\rho_b^j(A)\,\rho_c^k(A)\left\{\{\Phi_j,\Phi_k\},\Phi_i\right\}_{\Phi=0}-\rho_a^i(A)\,\rho_b^j(A)\,\rho_c^k(A)\left(\partial^m_p\{\Phi_j,\Phi_k\}\right)_{\Phi=0}\left\{\Phi_m,\Phi_i\right\}_{\Phi=0}\notag\\
&&+\mbox{cycl.}(abc)\,.\notag
\end{eqnarray}
The Jacobi identity implies that the first term on the right in (\ref{b3}) vanishes,
\begin{eqnarray}\label{b4}
&&\rho_a^i(A)\,\rho_b^j(A)\,\rho_c^k(A)\left\{\{\Phi_j,\Phi_k\},\Phi_i\right\}_{\Phi=0}+\mbox{cycl.}(abc)=\\
&&\rho_a^i(A)\,\rho_b^j(A)\,\rho_c^k(A)\left(\left\{\{\Phi_j,\Phi_k\},\Phi_i\right\}_{\Phi=0}+\left\{\{\Phi_i,\Phi_j\},\Phi_k\right\}_{\Phi=0}+\left\{\{\Phi_k,\Phi_i\},\Phi_j\right\}_{\Phi=0}\right)=0\,.\notag
\end{eqnarray}
Renaming the contracted indices we represent the second term on the right of (\ref{b3}) as,
\begin{eqnarray}\label{b5}
&&-\rho_a^i(A)\,\rho_b^j(A)\,\rho_c^k(A)\left(\partial^m_p\{\Phi_j,\Phi_k\}\right)_{\Phi=0}\left\{\Phi_m,\Phi_i\right\}_{\Phi=0}+\mbox{cycl.}(abc)=\\
&&-\rho_a^k(A)\,\rho_b^m(A)\,\rho_c^i(A)\left(\partial^j_p\{\Phi_m,\Phi_i\}\right)_{\Phi=0}\left\{\Phi_j,\Phi_k\right\}_{\Phi=0}+\mbox{cycl.}(abc)\,.\notag
\end{eqnarray}
Note that using the equation (\ref{rho}) one finds,
\begin{eqnarray}\label{b6}
&&-\rho_a^k(A)\,\rho_b^m(A)\,\rho_c^i(A)\left(\partial^j_p\{\Phi_m,\Phi_i\}\right)_{\Phi=0}=\\
&&-\rho_a^k(A)\,\rho_b^m(A)\,\rho_c^i(A)\left(\partial^j_p\{A_i,p_m\}\right)_{\Phi=0}+\rho_a^k(A)\,\rho_b^m(A)\,\rho_c^i(A)\left(\partial^j_p\{A_m,p_i\}\right)_{\Phi=0}=\notag\\
&&\rho_a^k(A)\,\{A_i,\rho_b^j(p)\}_{\Phi=0}\,\rho_c^i(A)-\rho_a^k(A)\,\rho_b^i(A)\,\{A_i,\rho_c^j(p)\}_{\Phi=0}\,.\notag
\end{eqnarray}
Taking into account (\ref{b6}) and the cyclic permutations we rewrite the r.h.s. of (\ref{b5}) as,
\begin{eqnarray}\label{b7}
&&\left(-\rho_a^i(A)\,\{A_i,\rho_b^j(p)\}_{\Phi=0}\,\rho_c^k(A)-\rho_a^i(A)\,\rho_b^j(A)\,\{A_i,\rho_c^k(p)\}_{\Phi=0}\right)\{\Phi_j,\Phi_k\}_{\Phi=0}+\\
&&\mbox{cycl.}(abc)\,.\notag
\end{eqnarray}

Remind that the expression (\ref{b7}) represents the modification of the last term in the r.h.s. of (\ref{b2}). Substituting (\ref{b7}) back into (\ref{b2}) we obtain for the r.h.s.,
\begin{eqnarray}\label{b8}
&&\rho_a^i(A)\left(\{\rho_b^j(A),\Phi_i\}_{\Phi=0}-\{A_i,\rho_b^j(p)\}_{\Phi=0}\right)\rho^k_c(A)\,\left\{\Phi_j,\Phi_k\right\}_{\Phi=0}+\\
&&\rho_a^i(A)\,\rho^j_b(A)\left(\{\rho_c^k(A),\Phi_i\}_{\Phi=0}-\{A_i,\rho_c^k(p)\}_{\Phi=0}\right)\rho^k_c(A)\,\left\{\Phi_j,\Phi_k\right\}_{\Phi=0}+\mbox{cycl.}(abc)\,.\notag
\end{eqnarray}
Let us calculate separately,
\begin{eqnarray}\label{b9}
&&\{\rho_b^j(A),\Phi_i\}_{\Phi=0}-\{A_i,\rho_b^j(p)\}_{\Phi=0}=\\
&&\partial^m_A\rho_b^j(A)\left(\{A_m,p_i\}_{\Phi=0}-\{A_i,p_m\}_{\Phi=0}-\{A_m,A_i\}\right)+\partial_m\rho_b^j(A)\,\gamma^m_i(A)=\notag\\
&&-\partial^m_A\rho_b^j(A)\,\{\Phi_m,\Phi_i\}_{\Phi=0}+\left(\partial_m\rho_b^j(x,p)\right)_{\Phi=0}\,\gamma^m_i(A)\,.\notag
\end{eqnarray}
Using (\ref{b9}) in (\ref{b8}) the latter becomes
\begin{eqnarray}\label{b10}
&&\left(\rho_a^i(A)\,\partial^m_A\rho_b^j(A)\,\rho^k_c(A)+\rho_a^i(A)\,\rho_b^j(A)\,\partial^m_A\rho^k_c(A)\right)\{\Phi_i,\Phi_m\}_{\Phi=0}\,\left\{\Phi_j,\Phi_k\right\}_{\Phi=0}+\\
&&\left(\rho_a^i(A)\,\gamma^m_i(A)\left(\partial_m\rho_b^j(x,p)\right)_{\Phi=0}\,\rho^k_c(A)+\rho_a^i(A)\,\rho_b^j(A)\,\gamma^m_i(A)\left(\partial_m\rho^k_c(x,p)\right)_{\Phi=0}\right)\left\{\Phi_j,\Phi_k\right\}_{\Phi=0}+\notag\\
&&\mbox{cycl.}(abc)\,.\notag
\end{eqnarray}
Renaming the contracted indices and using the cyclic permutations we may rewrite the first line of (\ref{b10}) as,
\begin{eqnarray}\label{b11}
\rho_a^i(A)\left(\partial^m_A\rho_b^j(A)-\partial^j_A\rho_b^m(A)\right)\rho^k_c(A)\,\{\Phi_i,\Phi_m\}_{\Phi=0}\,\left\{\Phi_j,\Phi_k\right\}_{\Phi=0}+\mbox{cycl.}(abc)\,.
\end{eqnarray}
Finally using (\ref{b11}) as well as the definitions (\ref{F}), (\ref{Lambda}) and (\ref{Kappa}) in (\ref{b10}) we rewrite the latter as,
\begin{eqnarray}\label{b12}
{\cal F}_{ad}\,\Lambda_b{}^{de}(A)\,{\cal F}_{ec}+\Gamma_{ab}{}^e(A)\,{\cal F}_{ec}-\Gamma_{cb}{}^e(A)\,{\cal F}_{ea}+\mbox{cycl.}(abc)\,.
\end{eqnarray}
The above expression is the right hand side of (\ref{b2}), which proves (\ref{bi1}). $\Box$

\section{Action principle and field equations}

Having in hand the covariant derivative (\ref{w6}) and the Poisson field strength (\ref{F}) now we are in the position to construct the gauge covariant Lagrangian, $\delta_f{\cal L}=\{{\cal L},f\}$, and gauge invariant action, $\delta_fS=0$. To do so, first one ought to introduce an appropriate integration mesure $\mu(x)$, which for any two Schwartz functions $f$ and $g$ should satisfy the condition,
\begin{equation}\label{measure}
\int\!d^Nx\,\mu(x)\,\{f,g\}=0\,,\qquad  \Leftrightarrow \qquad \partial_l\left(\mu(x)\,\Theta^{lk}(x)\right)=0\,.
\end{equation}
For the $su(2)$-like Poisson structure (\ref{su2}) the integration measure can be taken to be a constant $\mu(x)=1$. For $\kappa$-Minkowski Lie-Poisson bi-vector (\ref{kappa}) the measure is not constant. An explicit expression for $\mu(x)$ in this case was found in \cite{KKV}.

Let us discuss first the pure gauge sector without matter fields $\psi$. The gauge invariant action in this case reads \cite{KV20},
\begin{equation}\label{ag}
S_g=\int\!d^Nx\,\mu(x)\,{\cal L}_{g}\,,\qquad {\cal L}_{g}=-\frac14\,{\cal F}_{ab}\,{\cal F}^{ab}\,.
\end{equation}
To derive the corresponding Euler-Lagrange equations we will need the following relation.
{\proposition \label{peom}
Provided that the matrices $\gamma^i_l(A)$ and $\rho_a^l(A)$ are defined by the propositions \ref{pgt} and \ref{pcd} correspondingly, and the measure $\mu(x)$ satisfies (\ref{measure}), then the following relation holds true,
\begin{eqnarray}\label{tr1} 
\partial_i\left(\mu(x)\,\rho_a^l(A)\,\gamma^i_l(A)\right)+\mu(x)\left\{A_l,\rho_a^l(A)\right\}&=&-\mu(x)\,\Lambda_a{}^{cd}(A)\,{\cal F}_{cd}\\
&&+\left[\partial_i\left(\mu(x)\,\rho_a^l(p)\,\gamma^i_l(p)\right)\right]_{\Phi=0}\,,\notag
\end{eqnarray}
where the quantities $\Lambda_a{}^{cd}$ and ${\cal F}_{cd}$ were defined in the proposition \ref{cr}.}
{\proof Using the equation (\ref{rho}) we represent the left hand side as,
\begin{eqnarray*}
\left[\partial_i\left(\mu(x)\,\rho_a^l(p)\,\gamma^i_l(p)\right)\right]_{\Phi=0}+\partial^m_A\rho^l_a(A)\,\{\Phi_l,\Phi_m\}_{\Phi=0}\,,
\end{eqnarray*}
and use the definitions of $\Lambda_a{}^{cd}(A)$ and ${\cal F}_{cd}$ to complete the proof.} $\Box$

Using the relation (\ref{tr1}) one finds the Euler-Lagrange equations for (\ref{ag}),
\begin{eqnarray}
&&\mu\,\rho_b^k\,{\cal D}_a\left({\cal F}^{ab}\right)-\mu\,{\cal F}^{ab}\left(\rho_a^k\,\partial^m_A\rho_b^l+\rho_a^m\,\partial^l_A\rho_b^k+\rho_a^l\,\partial^k_A\rho_b^m\right)\{\Phi_l,\Phi_m\}_{\Phi=0}\\
&&+{\cal F}^{ab}\left[\partial_i\left(\mu\,\rho_a^l(p)\,\rho_b^k(p)\,\gamma^i_l(p)\right)\right]_{\Phi=0}=0\,.\notag
\end{eqnarray}
Multiplying by $\left(\rho^{-1}\right)^c_k$ we end up with Poisson gauge equations,
\begin{eqnarray}\label{ncm}
\mu\,{\cal D}_a\left({\cal F}^{ac}\right)+\frac{\mu}{2}\,{\cal F}^{cb}\,\Lambda_b{}^{de}\,{\cal F}_{de}-\mu\,{\cal F}^{db}\,\Lambda_b{}^{ce}\,{\cal F}_{de}+\left(\rho^{-1}\right)^c_k{\cal F}^{ab}\partial_i\left(\mu\,\rho_a^l\,\rho_b^k\,\gamma^i_l\right)=0\,.
\end{eqnarray}
These equations transform covariantly under the Poisson gauge transformation (\ref{APhi}) and reproduce the first pare of the Maxwell equations in the commutative limit $\alpha\to0$, if the measure $\mu(x)$ is constant.

As it was already mentioned, for the $su(2)$-like Poisson structure (\ref{su2}) the integration measure is constant $\mu(x)=1$. Since, for any linear Poisson structures the matrix $\rho_a^l(p)$ does not depend on coordinates, the last term on the left in (\ref{ncm}) disappears. Moreover, one may check that in this case,
\begin{equation}
{\cal F}^{ab}\left(\rho_a^k\,\partial^m_A\rho_b^l+\rho_a^m\,\partial^l_A\rho_b^k+\rho_a^l\,\partial^k_A\rho_b^m\right)\{\Phi_l,\Phi_m\}_{\Phi=0}\equiv0\,.
\end{equation}
Therefore the equations of motion of the $su(2)$-like Poisson gauge theory become,
\begin{eqnarray}\label{ncmsu2}
{\cal D}^{\varepsilon}_a\left({\cal F}^{ac}_\varepsilon\right)=0\,,
\end{eqnarray}
where the superscript $\varepsilon$ in ${\cal D}^{\varepsilon}_a$ and the subscript $\varepsilon$ in ${\cal F}^{ac}_\varepsilon$ just indicate that these objects correspond to the $su(2)$-structure.
For the $\kappa$-Minkowski case the measure is not constant, so the equations of motion will contain additional terms.

Now let us discuss the contribution from the matter fields. First of all let us observe that due proposition \ref{peom} one has,
\begin{eqnarray}
\int\!d^Nx\,\mu\,\psi\,{\cal D}_a(\psi)&=&-\int\!d^Nx\,\mu\,{\cal D}_a(\psi)\,\psi+\int\!d^Nx\,\mu\,\psi\,\Lambda_a{}^{cd}\,{\cal F}_{cd}\,\psi\\
&&+\int \!d^Nx\left[\partial_i\left(\mu(x)\,\rho_a^l(p)\,\gamma^i_l(p)\right)\right]_{\Phi=0}\psi\,\psi\,.\notag
\end{eqnarray}
The latter means that the operator ${\cal D}_a$ is not self-adjoint. Therefore to construct real gauge invariant action for the Fermionic matter fields $\psi$ we write,
\begin{equation}\label{am}
S_{int}=\int\!d^Nx\,\mu(x)\,{{\cal L}_{int}\,,\qquad\cal L}_{int}=\frac{i}{2}\left[\bar\psi\,\Gamma^a{\cal D}_a\psi-\left({\cal D}_a\bar\psi\right)\Gamma^a\,\psi\right]-m\,\bar\psi\,\psi\,,
\end{equation}
where $\Gamma^a$ are the gamma matrices. The corresponding Dirac equation reads,
\begin{equation}\label{ncpsi}
\left[i\,\mu\,\Gamma^a{\cal D}_a-\frac{i\mu}{2}\,\Gamma^a\,\Lambda_a{}^{cd}\,{\cal F}_{cd}-\frac{i}{2}\,\Gamma^a\,\partial_b\left(\mu\,\rho_a^l\,\gamma^b_l\right)-\mu\,m\right]\psi=0\,.
\end{equation}
To find the contribution of the matter fields to the Poisson Gauge equations one calculates,
\begin{eqnarray}
\frac{\delta S_{int}}{\delta A_k}=i\,\mu\,\rho_c^k\,\{\bar\psi\,\Gamma^c,\psi\}-\frac{i\mu}{2}\left(\partial^k_A\rho_c^l-\partial^l_A\rho_c^k\right)\left[\{\bar\psi,\Phi_l\}_{\Phi=0}\,\Gamma^c\,\psi-\bar\psi\,\Gamma^c\,\{\psi,\Phi_l\}_{\Phi=0}\right]\,
\end{eqnarray}
Multiplying it by $\left(\rho^{-1}\right)^a_k$ we obtain the right hand side (the current) of the eq. (\ref{ncm}),
\begin{equation}\label{nci}
J^a=-i\,\mu\,\{\bar\psi\,\Gamma^a,\psi\}+\frac{i\mu}{2}\,\Lambda_c{}^{ab}\left[\left({\cal D}_b\bar\psi\right)\Gamma^c\,\psi-\bar\psi\,\Gamma^c\,{\cal D}_b\psi\right]\,.
\end{equation}
In the commutative limit the interaction disappears, the field become electrically neutral as it was already discussed in the Sec. 4. 

\section{Conclusions and discussion}

Working in the framework of the symplectic embeddings we introduce the Poisson gauge transformations (\ref{APhi}) which close the algebra (\ref{ga}) obtained as the semi-classical limit of the full non-commutative gauge algebra (\ref{ga1}). The proposition \ref{pcd} determines the gauge covariant derivative of the matter field ${\cal D}_a(\psi)$. The commutator relation for the covariant derivatives (\ref{cr1}) defines the Poisson field strength (\ref{F}) which transforms covariantly under the gauge transformations (\ref{APhi}) and tends to the standard $U(1)$ field strength in the commutative limit. The corresponding Bianchi identity is set by the proposition \ref{bi}. The gauge invariant action, given by (\ref{ag}) for the pure gauge fields and by (\ref{am}) for the interaction part with the matter fields, is constructed in the Sec. 6 from the gauge covariant objects ${\cal D}_a(\psi)$ and ${\cal F}_{ab}$. From this action we derive the field equations (\ref{ncm}) for the gauge fields without interaction with the matter field, the Dirac equation for the matter field (\ref{ncpsi}), and the current term (\ref{nci}) describing the interaction between the gauge and the matter fields in the equations of motion. 

The symplectic embedding of given Poisson structure is not unique, the corresponding arbitrariness is described in the Sec. 2.1. So, for the same Poisson structure $\Theta^{ab}(x)$ one may obtain different Poisson gauge transformations and other subsequent structures. However the corresponding gauge systems are related by the Seiberg-Witten map described in the Sec. 3.1. If two different Poisson structures $\Theta^{ij}(x)$ and $\tilde\Theta^{ij}(\tilde x)$ are related by the invertible coordinate transformation, $\upsilon:\,x\to\tilde x$, then the corresponding Poisson gauge theories constructed following the proposed here scheme are related by the same coordinate transformation. 

For arbitrary given Poisson structure $\Theta^{ab}(x)$ all the structures appearing in the paper can be constructed up to arbitrary order in the deformation parameter $\alpha$. For some specific choices of the non-commutativity, like the rotationally invariant NC space (\ref{su2}), $\kappa$-Minkowski space (\ref{kappa}) or the spaces obtained from these two by the coordinate transformations we obtained explicit all-order in $\alpha$ expressions for all structures. It is interesting to work out different explicit examples of NC spaces, like e.g. the angular twist non-commutativity \cite{angular}. It would be also interesting to obtain an explicit solution of the equation (\ref{rho}) for any Lie-Poisson structure, $\Theta^{ab}(x)=f^{ab}_cx^c$, just like it was done in the Sec. 2.2 for the construction of the matrix $\gamma_a^l(x,p)$. 

In the abstract it is written that the model was designed to investigate the semi-classical features of the complete non-commutative gauge theory. So, it would be interesting to understand what kind of physics may stay behind the Poisson gauge field equations obtained in the Sec. 6. It would be interesting to investigate in particular the corresponding energy-momenta dispersion relations.

The most non-trivial and at the same time interesting part of \cite{KS21} is related to the construction of the almost-Poisson gauge algebra. There are certain configurations in string and $M$-theory like the non-geometric backgrounds \cite{NA1}-\cite{NA4} yielding not only non-commutative but also non-associative deformations of space-time \cite{Szabo2}-\cite{KS17}. In this case the corresponding algebra of brackets (\ref{PB}) does not satisfy the Jacobi identity. But even so it is possible to construct the Lie algebra of gauge symmetries defined on such non-associative spaces. The key difference with the Poisson gauge algebra (\ref{ga}) consists in the fact that the commutator of two gauge transformations is again a gauge transformation, however with a field-dependent gauge parameter. It is challenging to generalize the constructions obtained in the present research to the case of almost-Poisson deformations and thus to construct the semi-classical limit of non-associative gauge theory and non-associative gravity.

\subsection*{Acknowledgments}

I am grateful to Dima Vassilevich for fruitful discussions. I appreciate the support from the Funda\c c\~ao de Amparo a Pesquisa do Estado de S\~ao Paulo (FAPESP, Brazil).

\section*{Appendix: Useful formulae}

Let us first calculate,
\begin{eqnarray}\label{r1}
&&\{f(x),\{g(x),p_a\}\}_{\Phi=0}-\{f(x),\{g(x),p_a\}_{\Phi=0}\}_{\Phi=0}=\\
&&\{f(x),\gamma^i_a(x,p)\,\partial_ig(x)\}_{\Phi=0}-\{f(x),\gamma^i_a(x,A(x))\,\partial_ig(x)\}_{\Phi=0}=\notag\\
&&\left(\partial^b_p\{g(x),p_a\}\right)_{\Phi=0}\left(\{f(x),p_b\}_{\Phi=0}-\{f(x),A_b(x)\}_{\Phi=0}\right)=\notag\\
&&\left(\partial^b_p\{g(x),p_a\}\right)_{\Phi=0}\{f(x),\Phi_b\}_{\Phi=0}\,.\notag
\end{eqnarray}
Observe that by (\ref{PB}) and (\ref{PB1}) the Poisson bracket of two functions $f(x)$ and $g(x)$ of coordinates only does not depend on $p$-variables, so,
\begin{eqnarray}\label{r2}
\{f(x),g(x)\}_{\Phi=0}=\{f(x),g(x)\}\,.
\end{eqnarray}
The combination of (\ref{r1}) and (\ref{r2}) implies,
\begin{equation}\label{r3}
\left\{f(x),\{g(x),\Phi_a\}\right\}_{\Phi=0}-\left\{f(x),\{g(x),\Phi_a\}_{\Phi=0}\right\}_{\Phi=0}=\left(\partial^d_p\{g(x),\Phi_a\}\right)_{\Phi=0}\{f(x),\Phi_d\}_{\Phi=0}\,.
\end{equation}

One may also check that,
\begin{eqnarray}\label{r5}
&&\{\{f(x),\Phi_b\},\Phi_a\}_{\Phi=0}-\{\{f(x),\Phi_b)\}_{\Phi=0},\Phi_a\}_{\Phi=0}=\\
&&\left(\partial^d_p\{f(x),\Phi_b)\}\right)_{\Phi=0}\{\Phi_d,\Phi_a\}_{\Phi=0}\,,\notag
\end{eqnarray}
and,
\begin{eqnarray}\label{r6}
&&\{\{\Phi_c,\Phi_b\},\Phi_a\}_{\Phi=0}-\{\{\Phi_c,\Phi_b)\}_{\Phi=0},\Phi_a\}_{\Phi=0}=\\
&&\left(\partial^d_p\{\Phi_c,\Phi_b)\}\right)_{\Phi=0}\{\Phi_d,\Phi_a\}_{\Phi=0}\,.\notag
\end{eqnarray}

\end{document}